\documentclass[10pt,letterpaper,twocolumn]{article} %% two column, final layout

\usepackage[tablesfirst,notablist,nomarkers]{endfloat} %% float figs. to back

\usepackage{ol}
\usepackage{hyperref}
\usepackage{amsmath}
\usepackage{bm}% bold math
\usepackage{amssymb}

\begin{document}

\twocolumn[ %% activate for two-column option

\title{Angular spectrum of quantized light beams}

\author{A. Aiello, J. Visser, G. Nienhuis, and J. P. Woerdman}

\address{Huygens Laboratory, Leiden University\\
P.O.\ Box 9504, 2300 RA Leiden, The Netherlands}

% Do not use \email or \homepage here. E-mail and URL can be given just before references.

\begin{abstract}
We introduce a generalized angular spectrum representation for
quantized light beams. By using our formalism, we are able to
derive simple expressions for the electromagnetic vector potential
operator in the case of: {a)} time-independent paraxial fields,
{b)} time-dependent paraxial fields, and {c)} non-paraxial fields.
For the first case, the well known paraxial results are fully
recovered.
\end{abstract}

\ocis{000.1600, 270.0270.}

 ] %% activate for two-column option

Propagation of nonclassical states of the electromagnetic field is
an issue of growing interest in quantum optics for both
fundamental and technological purposes \cite{Kolo99}. Consider,
for instance, the relevance of propagation of entangled photons to
quantum cryptographic systems \cite{Gisin02}.

The purpose of the present Letter is to provide a novel perfectly
general formalism for the representation of quantized light beams
 which can be used in \emph{any} regime of propagation. Such objective
is achieved by using a dispersion relation that some of us have
recently introduced in Ref. \cite{Aiello_05para}, and a
generalized
 angular spectrum representation for the field
operators \cite{Visser_05}.
 The usefulness of our approach \cite{Calvo} becomes manifest whenever one deals
with quantum systems for which both paraxial and non-paraxial
regimes of propagation may be relevant as, e.g., down-converted
photon pairs \cite{MandelBook}.

Consider the plane-wave expansion of  the positive-frequency part
of the electromagnetic vector potential operator
$\hat{\mathbf{A}}(\mathbf{r},t) =
\hat{\mathbf{A}}^{(+)}(\mathbf{r},t) + \mathrm{H.c.}$ in the
Coulomb gauge \cite{LoudonBook}
\begin{equation}\label{a10}
\begin{array}{lcl}
  \displaystyle{ \hat{\mathbf{A}}^{(+)}(\mathbf{r},t)   } & = & \displaystyle{ \int \mathrm{d}^3 \mathbf{k}
  \left( \frac{\hbar}{16 \pi^3 \varepsilon_0 c |\mathbf{k}|} \right)^{1/2}}\\
   & & \displaystyle{\times  \sum_{\lambda = 1}^2 {\bm \epsilon}^{(\lambda)}(\mathbf{k})
  \hat{a}_\lambda(\mathbf{k}) \exp \left( \mathrm{i} \mathbf{k} \cdot \mathbf{r} - \mathrm{i} c |\mathbf{k}| t \right)} .\\
\end{array}
\end{equation}
Since we want to describe fields propagating mainly along the $z$
axis, we find it convenient to define  $k_z = s \zeta$, where $ s
\equiv \mathrm{sign}( k_z )= \pm 1$, and
 $\zeta \geq 0$. Then we can rewrite Eq. (\ref{a10}) as
\begin{equation}\label{a30}
\begin{array}{l}
  \displaystyle{ \hat{\mathbf{A}}^{(+)}(\mathbf{r},t)  = }  \displaystyle{  \sum_{s = \pm 1} \int
  \mathrm{d}k_x  \mathrm{d}k_y  \int_0^{\infty} \mathrm{d} \zeta
  \left( \frac{\hbar}{16 \pi^3 \varepsilon_0 c |\mathbf{k}_s|} \right)^{1/2}}\\
    \displaystyle{\; \; \; \; \; \; \; \; \; \; \; \; \; \; \; \; \, \times
   \sum_{\lambda = 1}^2 {\bm \epsilon}^{(\lambda)}(\mathbf{k}_s)
  \hat{a}_\lambda(\mathbf{k}_s) \exp \left( \mathrm{i} \mathbf{k}_s \cdot \mathbf{r} - \mathrm{i} c |\mathbf{k}_s| t \right)} ,\\
\end{array}
\end{equation}
where we have defined $\mathbf{k}_s = (k_x,k_y, s \zeta)$. The
field annihilation and creation operators satisfy the canonical
commutation relations \cite{LoudonBook}
\begin{equation}\label{a60}
[ \hat{a}_\lambda(\mathbf{k}_s),
\hat{a}_{\lambda'}^\dagger(\mathbf{k}'_{s'})] = \delta_{\lambda
\lambda'}  \delta_{s s'} \delta^{(2)}(\mathbf{q} - \mathbf{q}')
\delta(\zeta - \zeta').
\end{equation}

Let $\omega \geq 0$ be an arbitrary frequency; at a later point in
this Letter we shall identify $\omega$ with the carrier frequency
of a paraxial field. We perform a  change of variables
$\{k_x,k_y,\zeta \} \rightarrow \{ q_x,q_y ,\omega \}$ such that
\begin{equation}\label{a40}
k_x = q_x, \quad
k_y = q_y, \quad
\zeta = f(\mathbf{q}, \omega),
\end{equation}
where $f(\mathbf{q}, \omega) \geq 0$ is an (almost) arbitrary
function to be determined. For  reasons that will be soon clear,
we require $f(\mathbf{q}, \omega)$ to increase monotonically for
increasing $\omega$ in the domain
\begin{equation}\label{a50}
\mathcal{I}_\omega(f,\mathbf{q}) = \left\{ \omega \in \mathbb{R}^+
: f(\mathbf{q},\omega) \geq 0 \right\}.
\end{equation}
This condition implies  that $\mathrm{d} f(\mathbf{q},\omega) /
\mathrm{d} \omega > 0$ for $\omega \in
\mathcal{I}_\omega(f,\mathbf{q})$.
Therefore, in such domain, we can write
\begin{equation}\label{a70}
\begin{array}{l}
\displaystyle{\delta^{(2)}(\mathbf{q} - \mathbf{q}') \delta(\zeta
- \zeta')}  =  \displaystyle{\delta^{(2)}(\mathbf{q} -
\mathbf{q}')
\delta \left[ f(\mathbf{q},\omega) - f(\mathbf{q},\omega') \right]}\\\\
  \; \; \; \; \;  \; \; \; \; \; \; \; \; \; \;  \; \; \; \; \; \;  \; \; \; \; \; \; \; \; \; \;  \; \;
    \displaystyle{= \delta^{(2)}(\mathbf{q} - \mathbf{q}')\frac{\delta(\omega - \omega')}
    {\mathrm{d} f(\mathbf{q},\omega) / \mathrm{d}
\omega}}.\\
\end{array}
\end{equation}
If we substitute Eq. (\ref{a70}) into  Eq. (\ref{a60}), we obtain
\begin{equation}\label{a80}
[ \hat{a}_\lambda(\mathbf{k}_s),
\hat{a}_{\lambda'}^\dagger(\mathbf{k}'_{s'})] = \delta_{\lambda
\lambda'}  \delta_{s s'} \delta^{(2)}(\mathbf{q} -
\mathbf{q}')\frac{\delta(\omega - \omega')}{\mathrm{d}
f(\mathbf{q},\omega) / \mathrm{d} \omega}.
\end{equation}
Equation  (\ref{a80}) suggests  the introduction of the
``angular-spectrum'' field operators $\hat{a}_{\lambda
s}(\mathbf{q}, \omega)$ defined as \cite{Visser_05}
\begin{equation}\label{a90}
\hat{a}_{\lambda s}(\mathbf{q}, \omega) \equiv
\hat{a}_\lambda(\mathbf{k}_s) \sqrt{\mathrm{d}
f(\mathbf{q},\omega) / \mathrm{d} \omega} .
\end{equation}
By using  Eq. (\ref{a90}) and  Eq. (\ref{a80}) it is easy to see
that
\begin{equation}\label{a100}
[ \hat{a}_{\lambda s}(\mathbf{q}, \omega), \hat{a}_{\lambda'
s'}^\dagger(\mathbf{q}', \omega')] = \delta_{\lambda \lambda'}
\delta_{s s'} \delta^{(2)}(\mathbf{q} - \mathbf{q}')\delta(\omega
- \omega').
\end{equation}
Equation (\ref{a100}) is the first main result of this Letter; it
worthwhile to note that  it is \emph{exact}. No approximations
were made to obtain it.

The condition $f(\mathbf{q},\omega) \geq 0$ defines a volume
$\mathcal{V}_{\mathbf{q}, \omega}(f)$ in the half-space
$\mathbb{R}^2 \times \mathbb{R}^+$ spanned by $\{q_x,q_y,\omega
\}$. This volume is bounded by the surface
$\mathcal{S}_{\mathbf{q}, \omega}(f)=
\partial\mathcal{V}_{\mathbf{q}, \omega}(f)$ defined by the equation $f(\mathbf{q},\omega) = 0$.
If we define the $2$-dimensional domain
$\mathcal{C}_\mathbf{q}(f,\omega) = \left\{ (q_x,q_y)\in
\mathbb{R}^2 : f(\mathbf{q},\omega) \geq 0 \right\}$
 then, from Eq. (\ref{a50}) it readily follows
\begin{equation}\label{a120}
\int_{\mathbb{R}^2} \mathrm{d}^2 \mathbf{q}
\int_{\mathcal{I}_\omega(f,\mathbf{q})} \mathrm{d} \omega =
\int_{\mathbb{R}^+} \mathrm{d} \omega
\int_{\mathcal{C}_\mathbf{q}(f,\omega)} \mathrm{d}^2 \mathbf{q}.
\end{equation}
We use this equality to rewrite Eq. (\ref{a30})  immediately in
the new variables $\{q_x,q_y, \omega \}$ as
\begin{equation}\label{a130}
\begin{array}{l}
  \displaystyle{ \hat{\mathbf{A}}^{(+)}(\mathbf{r},t)  = }  \displaystyle{  \sum_{s = \pm 1}
  \int_0^{\infty} \mathrm{d} \omega \int_{\mathcal{C}_\mathbf{q}(f,\omega)}
\! \! \! \!   \mathrm{d}^2 \mathbf{q}
  \left( \frac{\hbar/\sqrt{q^2 + f^2}}{16 \pi^3 \varepsilon_0 c } \right)^{1/2}}\\
\; \; \; \; \;  \; \; \; \; \; \; \; \; \; \;  \; \,
\displaystyle{\times \sum_{\lambda = 1}^2
   {\bm \epsilon}^{(\lambda)}(\mathbf{q},sf) \hat{a}_{\lambda s}(\mathbf{q},\omega)\sqrt{\mathrm{d} f /
\mathrm{d} \omega}}\\
\; \; \; \; \;  \; \; \; \; \; \; \; \; \; \;  \; \,
\displaystyle{ \times  \exp \left( \mathrm{i} \mathbf{q} \cdot \mathbf{x} +
 \mathrm{i} s f z  - \mathrm{i}  t c \sqrt{q^2 + f^2}  \right)} ,\\
\end{array}
\end{equation}
where $q^2 = q_x^2 + q_y^2$, $f = f(\mathbf{q},\omega)$, and
$\mathbf{x} = (x,y)$. Note that the square root of the Jacobian $J
= \mathrm{d} f / \mathrm{d} \omega$ was used to pass from the
original operators $\hat{a}_{\lambda}(\mathbf{k}_s)$ to the
angular-spectrum operators $\hat{a}_{\lambda
s}(\mathbf{q},\omega)$. Since we want to develop a formalism
suitable for both non-paraxial and paraxial light beams, we
rewrite Eq. (\ref{a130}) as
\begin{equation}\label{a150}
  \displaystyle{ \hat{\mathbf{A}}^{(+)}(\mathbf{r},t)  =  \sum_{s = \pm 1}
  \int_0^{\infty} \mathrm{d} \omega \,
   e^{-\mathrm{i} \omega (t - s z/c)} \, {\hat{\bm \Psi}_s} (\mathbf{r},t;
\omega)},
\end{equation}
so that $\omega$ determines the plane carrier wave, and ${\hat{\bm
\Psi}_s} (\mathbf{r},t; \omega)$ is the \emph{envelope} field
which, at this stage, is not required to be spatially and
temporally slowly varying:
\begin{equation}\label{a160}
\begin{array}{lcl}
  \displaystyle{ {\hat{\bm \Psi}_s} (\mathbf{r},t;
\omega)   }  & =&\displaystyle{
\int_{\mathcal{C}_\mathbf{q}(f,\omega)}
  \mathrm{d}^2 \mathbf{q}
  \left( \frac{\hbar \, {\mathrm{d} f /
\mathrm{d} \omega}}{16 \pi^3 \varepsilon_0 c \sqrt{q^2 + f^2}} \right)^{1/2}}\\
 &  & \displaystyle{\times  \sum_{\lambda = 1}^2
   {\bm
\epsilon}^{(\lambda)}_s(\mathbf{q},\omega) \hat{a}_{\lambda
s}(\mathbf{q},\omega)}\\
 &  &
\displaystyle{ \times  \exp \left[ \mathrm{i} \mathbf{q} \cdot
\mathbf{x} + \mathrm{i} z s(f- \omega/c) \right]}
\\
 &  &
\displaystyle{ \times  \exp \left[  - \mathrm{i} t  \left(
c\sqrt{q^2 + f^2} -\omega \right) \right]}
 .\\
\end{array}
\end{equation}
Moreover, since $\mathbf{k}_s =  \hat{\mathbf{k}}_s \sqrt{q^2 +
f^2}$, where $\hat{\mathbf{k}}_s = (q \hat{\mathbf{q}} + s f
\hat{\mathbf{z}} )/\sqrt{q^2+f^2}$, and $\hat{\mathbf{q}} =
\mathbf{q}/q$; we have defined $ {\bm
\epsilon}^{(\lambda)}_s(\mathbf{q},\omega) \equiv {\bm
\epsilon}^{(\lambda)}(\mathbf{q},sf)$, where $ {\bm
\epsilon}^{(2)}_s(\mathbf{q},\omega) = s \hat{\mathbf{z}} \times
\hat{\mathbf{q}}$, and
\begin{equation}\label{a165}
{\bm \epsilon}^{(1)}_s(\mathbf{q},\omega) = (f \hat{\mathbf{q}} -
s q \hat{\mathbf{z}} )/\sqrt{q^2 + f^2}.
\end{equation}
Until now, we furnished expressions for the field operators in the
angular spectrum representation, but not for the energy, the
momentum, etc. However, closed expressions for these physical
quantities can be easily found  by noting that the
 product
\begin{equation}\label{a290}
 \hat{a}_{\lambda}^\dagger(\mathbf{k}_s)
\hat{a}_{\lambda}(\mathbf{k}_s) \, \mathrm{d} \zeta =
 \hat{a}_{\lambda s}^\dagger(\mathbf{q},
\omega) \hat{a}_{\lambda s}(\mathbf{q}, \omega)\,\mathrm{d}
\omega,
\end{equation}
is invariant with respect to the change of variables Eq.
(\ref{a40}). Then, for example, starting from the well known
expression for the Hamiltonian operator of the electromagnetic
field (see, e.g., Ref. \cite{LoudonBook}), after a straightforward
calculation one obtain
\begin{equation}\label{a310}
\begin{array}{l}

  \displaystyle{ \hat{H}} =  \displaystyle{ \frac{1}{2} \sum_{s = \pm 1}
  \int_0^{\infty} \mathrm{d} \omega \,\int_{\mathcal{C}_\mathbf{q}(f,\omega)}
  \mathrm{d}^2 \mathbf{q} \, \hbar c \sqrt{q^2 + f^2(\mathbf{q},\omega)}
    }
\\
  \; \; \; \; \, \displaystyle{\times \sum_{\lambda = 1}^2
\left[ \hat{a}_{\lambda s}^\dagger (\mathbf{q},\omega)
\hat{a}_{\lambda s}(\mathbf{q},\omega)  + \hat{a}_{\lambda s}
(\mathbf{q},\omega) \hat{a}_{\lambda
s}^\dagger(\mathbf{q},\omega)\right].
  }\\
\end{array}
\end{equation}
Similar calculations can be easily done for the other relevant
quantities. Equation (\ref{a310}) shows that, as expected for an
arbitrary field, the frequency $\omega$ of the carrier plane wave
 is not equal to the frequency
$c|\mathbf{k}|=c\sqrt{q^2 + f^2(\mathbf{q},\omega)}$ of the
plane-wave mode $\exp(\mathrm{i} \mathbf{k} \cdot \mathbf{r})$.
However, as we shall see later, $c|\mathbf{k}|$ reduces to
$\omega$ in the paraxial limit.

 At this point the function
$f(\mathbf{q}, \omega)$ is still undetermined, therefore  we can
exploit this freedom by imposing some constraints on the envelope
field ${\hat{\bm \Psi}_s} (\mathbf{r},t; \omega)$ which is, until
now, perfectly general.
In particular, we want to find an expression for the envelope
field in which  the  Fresnel propagator \cite{MandelBook} plays a
role even beyond the paraxial regime.
To this end, we proceed as in Ref. \cite{Aiello_05para} and we
require ${\hat{\bm \Psi}_s} (\mathbf{r},t = 0; \omega) \equiv
{\hat{\bm \Psi}_s} (\mathbf{r}; \omega)$ to satisfy the
time-independent paraxial equation:
\begin{equation}\label{a170}
\frac{\partial^2 {\hat{\bm \Psi}_s}(\mathbf{r}; \omega)}{\partial
x^2} + \frac{\partial^2 {\hat{\bm \Psi}_s}(\mathbf{r};
\omega)}{\partial y^2} + 2 \mathrm{i} s \frac{\omega}{c}
\frac{\partial {\hat{\bm \Psi}_s}(\mathbf{r}; \omega)}{\partial z}
= 0.
\end{equation}
In this way we obtain an expression for
$\hat{\mathbf{A}}^{(+)}(\mathbf{r},t)$ which is an \emph{exact}
solution of the full d'Alembert equation for any time $t>0$ and
its corresponding envelope field ${\hat{\bm \Psi}_s} (\mathbf{r},t
= 0; \omega)$ satisfies the time-independent paraxial wave
equation at $t=0$, as initial condition.
If we substitute from Eq. (\ref{a160}) the plane wave $\exp \left[
\mathrm{i} \mathbf{q} \cdot \mathbf{x} + \mathrm{i} z s(f-
\omega/c) \right]$ into Eq. (\ref{a170}), we easily find
\begin{equation}\label{a190}
f(\mathbf{q},\omega) = \frac{\omega}{c} \left( 1 - \frac{q^2
c^2}{2 \omega^2} \right).
\end{equation}

For $\mathbf{q} \in \mathcal{C}_{\mathbf{q}}(f,\omega)$ this
function satisfies all our requirements: it is positive and $
 {\mathrm{d}
f(\mathbf{q},\omega)}/{\mathrm{d} \omega} =
 \left( 1 + \vartheta^2 \right)/c >0$ where we have defined \cite{NotePara}
 $\vartheta \equiv {q c}/(\sqrt{2} \omega)$
.
It is easy to see that the plane-wave frequency
 $c |\mathbf{k}| = \omega (1 + \vartheta^4)^{1/2}$ reduces to
 $\omega$ in the paraxial limit $\vartheta \ll 1$. Finally, a
 closed expression for the field operator
 $\hat{\mathbf{A}}^{(+)}(\mathbf{r},t)$ can be given:
\begin{equation}\label{a300}
\begin{array}{ll}
 &  \! \! \! \! \! \! \! \! \! \! \! \! \! \! \! \!
  \displaystyle{ \hat{\mathbf{A}}^{(+)}} \displaystyle{(\mathbf{r},t) } =  \displaystyle{  \sum_{s = \pm 1}
  \int_0^{\infty} \mathrm{d} \omega \,
   e^{-\mathrm{i} \omega (t - s z/c)} \, }
\\
  & \displaystyle{\times
\int_{\mathcal{C}_\mathbf{q}(f,\omega)}
  \mathrm{d}^2 \mathbf{q}
  \left( \frac{\hbar(1 + \vartheta^2)}{16 \pi^3 \varepsilon_0 c \omega \sqrt{1 + \vartheta^4} } \right)^{1/2}
  }\\
   & \displaystyle{\times  \sum_{\lambda = 1}^2
   {\bm \epsilon}^{(\lambda)}_s(\mathbf{q},\omega) \hat{a}_{\lambda s}(\mathbf{q},\omega)
   \exp \left( \mathrm{i} \mathbf{q} \cdot
\mathbf{x} - \mathrm{i}  s \frac{q^2 c }{2 \omega} z \right)}\\
   &
\displaystyle{ \times  \exp \left[  - \mathrm{i} \omega t  \left(
\sqrt{1 + \vartheta^4} -1 \right) \right]}
 .\\
\end{array}
\end{equation}
Equation (\ref{a300}) is the second  main result of this Letter.
It is easy to recognize in the exponential function in the third
row, the sought Fresnel propagator in momentum space. The spatial
behavior of the envelope field is entirely governed by this term.
It worth to note that Eq. (\ref{a300}) is \emph{exact}, that is it
has been obtained without any approximation and, therefore, it
holds for both non-paraxial ($\vartheta \lesssim 1$) and paraxial
($\vartheta \ll 1$) beams. In the latter case, the slowly varying
term $ \exp \left[  - \mathrm{i} \omega t \left( \sqrt{1 +
\vartheta^4} -1 \right) \right]$ shows that the envelope field
${\hat{\bm \Psi}_s} (\mathbf{r},t; \omega)$ cannot be strictly
monochromatic for any $t >0$.

In the remaining part of this Letter, we  give two different
examples of the application of our theory in order to illustrate
its  generality. As a first example, let us generalize the
previous case and require ${\hat{\bm \Psi}_s} (\mathbf{r},t;
\omega)$  to satisfy the \emph{time-dependent} paraxial wave
equation, for any $t$ \cite{Deutsch91a}:
\begin{equation}\label{a180}
\frac{\partial^2 {\hat{\bm \Psi}_s}}{\partial x^2} +
\frac{\partial^2 {\hat{\bm \Psi}_s}}{\partial y^2} + 2 \mathrm{i}
s \frac{\omega}{c} \frac{\partial {\hat{\bm \Psi}_s}}{\partial z}
+ 2 \mathrm{i} \frac{\omega}{c^2} \frac{\partial {\hat{\bm
\Psi}_s}}{\partial t} = 0,
\end{equation}
where ${\hat{\bm \Psi}_s}  \equiv {\hat{\bm \Psi}_s}
(\mathbf{r},t; \omega)$ for short. If we substitute from Eq.
(\ref{a160}) the relevant term $\exp [ \mathrm{i} \mathbf{q} \cdot
\mathbf{x} + \mathrm{i} z s(f- \omega/c) ] \times \exp [ -
\mathrm{i} t  ( c\sqrt{q^2 + f^2} -\omega ) ]$ into Eq.
(\ref{a180}), we obtain a new dispersion relation
\begin{equation}\label{a225}
f(\mathbf{q},\omega) = \frac{\omega}{c} \left( 1 - \frac{q^2
c^2}{4 \omega^2} \right).
\end{equation}
Once again, for $\mathbf{q} \in
\mathcal{C}_{\mathbf{q}}(f,\omega)$ this function satisfies all
our requirements: it is positive and $ {\mathrm{d}
f(\mathbf{q},\omega)}/{\mathrm{d} \omega} =
 \left( 1 + \eta^2 \right)/c >0$,
where  $\eta \equiv {q c}/(2 \omega)$.
It is easy to see that the plane-wave frequency $c |\mathbf{k}|$
becomes  $ c |\mathbf{k}| = \omega (1 + \eta^2)$.
Also for this case a
 closed expression for the field operator
 $\hat{\mathbf{A}}^{(+)}(\mathbf{r},t)$ can be given:
\begin{equation}\label{a240}
\begin{array}{ll}
 & \! \! \! \! \! \!
  \displaystyle{ \hat{\mathbf{A}}^{(+)}} \displaystyle{(\mathbf{r},t) } =  \displaystyle{  \sum_{s = \pm 1}
  \int_0^{\infty} \mathrm{d} \omega \,
   e^{-\mathrm{i} \omega (t - s z/c)} \, }
\\
  & \displaystyle{\times
\int_{\mathcal{C}_\mathbf{q}(f,\omega)}
  \mathrm{d}^2 \mathbf{q}
  \left( \frac{\hbar}{16 \pi^3 \varepsilon_0 c \omega } \right)^{1/2}
  }\\
   & \displaystyle{\times  \sum_{\lambda = 1}^2
   {\bm \epsilon}^{(\lambda)}_s(\mathbf{q},\omega) \hat{a}_{\lambda s}(\mathbf{q},\omega)
   \exp \left[ \mathrm{i} \mathbf{q} \cdot
\mathbf{x} - \mathrm{i}   \frac{q^2 c }{4 \omega}(sz + c t)\right]}.\\
\end{array}
\end{equation}
This expression is quite simpler than Eq. (\ref{a300}). However,
its exponential part (in the last row), differ by a factor of
$1/2$ from the Fresnel propagator expression. Once again, we
stress that Eq. (\ref{a240}) is exact, no approximation were made.

As a last example of application of our formalism, we choose to
determine the function $f(\mathbf{q},\omega)$ by requiring
$\omega$ to coincide with the plane-wave frequency
\cite{Visser_05} $c |\mathbf{k}|: \, \omega = c |\mathbf{k}|$. It
is easy to see that in this case we have
\begin{equation}\label{a250}
f(\mathbf{q},\omega) = \frac{\omega}{c} \left( 1 - \frac{q^2
c^2}{\omega^2} \right)^{1/2} \simeq \frac{\omega}{c} \left( 1 -
\frac{q^2 c^2}{2 \omega^2} \right),
\end{equation}
where the last approximate equality holds in the paraxial limit
$qc/\omega \ll 1$.  For $\mathbf{q} \in
\mathcal{C}_{\mathbf{q}}(f,\omega)$ this function is positive and
${\mathrm{d} f(\mathbf{q},\omega)}/{\mathrm{d} \omega} = 1/ ( c
\sqrt{1 - (qc/\omega)^2 }) >0$, therefore all our requirements are
fulfilled. As expected,  in the paraxial limit Eq. (\ref{a250})
coincides with Eq. (\ref{a190}). Since by definition $\zeta =
|k_z| = (\omega/c)|\cos \theta|$, it follows that $|\cos \theta| =
f(\mathbf{q},\omega) c/\omega$, and we can write
\begin{equation}\label{a270}
\begin{array}{ll}
  \displaystyle{ \hat{\mathbf{A}}^{(+)}} \displaystyle{(\mathbf{r},t) } =  \displaystyle{  \sum_{s = \pm 1}
  \int_0^{\infty} \mathrm{d} \omega \,
   e^{-\mathrm{i} \omega (t - s z/c)} \, }
\\
   \displaystyle{ \; \;  \; \times
\int_{\mathcal{C}_\mathbf{q}(f,\omega)}
  \mathrm{d}^2 \mathbf{q}
  \left( \frac{\hbar/|\cos \theta|}{16 \pi^3 \varepsilon_0 c \omega } \right)^{1/2}
\sum_{\lambda = 1}^2
   {\bm \epsilon}^{(\lambda)}_s(\mathbf{q},\omega) \hat{a}_{\lambda s}(\mathbf{q},\omega)
  }\\
    \displaystyle{\; \;  \;  \times
   \exp \left[ \mathrm{i} \mathbf{q} \cdot
\mathbf{x} - \mathrm{i} sz\omega (1-|\cos \theta|)/c\right]}.\\
\end{array}
\end{equation}
Equation (\ref{a270}) is our last result. It gives an exact
expression for the electromagnetic potential vector operator of a
generic, non-paraxial light beam, in the angular spectrum
representation. By expanding in Taylor series the $|\cos \theta|$
term around $\theta =0$, it is easy to see that Eq. (\ref{a270})
reduces to the well know classical paraxial expression (with the
quantum operators $\hat{a}_{\lambda s}(\mathbf{q},\omega)$
substituted by the corresponding classical amplitudes). Moreover,
at the lowest order in $\theta$, it coincides with Eq.
(\ref{a300}) calculated at the lowest order in $\vartheta$.

In conclusion, in this Letter we presented a novel formalism for
the representation of arbitrary quantized light beams. First, we
introduced an angular spectrum representation for the field
annihilation and creation operators. Then, we used our formalism
to derive an exact expression for the ``paraxial-like'' envelope
field of a light beam. Finally, we illustrated the generality of
our theory, by applying it to the description of time-dependent,
paraxial and non-paraxial, light beams.
It worth to note that, although our formalism is fully quantum,
\emph{all} the previous results can be straightforwardly extended
to classical fields just by replacing the quantum operators
$\hat{a}_{\lambda s}(\mathbf{q},\omega)$ with the corresponding
classical amplitudes $a_{\lambda s}(\mathbf{q},\omega)$.

%\begin{acknowledgments}
We acknowledge support from the EU under the IST-ATESIT contract.
This project is also supported by FOM.
%\end{acknowledgments}
%

\end{document}